\documentclass{ws-procs961x669}            

\usepackage{tikz}

\newcommand{\R}{\mathbb{R}}
\newcommand{\C}{\mathbb{C}}
\newcommand{\Z}{\mathbb{Z}}
\newcommand{\bbZ}{\mathbb{Z}}
\newcommand{\fa}{\mathfrak{a}}

\newcommand{\bF}{\mathbf{F}}
\newcommand{\bS}{\mathbf{S}}
\newcommand{\bL}{\mathbf{L}}
\newcommand{\bg}{\mathbf{g}}
\newcommand{\fh}{\mathfrak{h}}
\newcommand{\bAKN}{\mathbf{A}_{\textsc{KN}}}

\newcommand{\ga}{\gamma}

\newcommand{\Om}{\Omega}
\newcommand{\la}{\lambda}
\newcommand{\pd}{\partial}
\newcommand{\ka}{\kappa}
\newcommand{\mc}{\mathcal}
\newcommand{\si}{\sigma}
\newcommand{\al}{\alpha}

\newcommand{\cC}{\mathcal{C}}

\newcommand{\fD}{\mathfrak{D}}

\newcommand{\Sset}{\mathbb{S}}

\newcommand{\De}{\varpi}
\newcommand{\half}{\frac{1}{2}}
\newcommand{\sgn}{\mbox{sgn}}

\newcommand{\bna}{\begin{eqnarray}}
\newcommand{\ena}{\end{eqnarray}}

\newcommand\beq{\begin{equation}}
\newcommand\eeq{\end{equation}}
\newcommand\ben{\begin{enumerate}}
\newcommand\een{\end{enumerate}}
\newcommand\bit{\begin{itemize}}
\newcommand\eit{\end{itemize}}

\begin{document}
\title{The point spectrum of the Dirac Hamiltonian \\on the zero-gravity Kerr-Newman spacetime}

\author{M. Kiessling, E. Ling, and A. S. Tahvildar-Zadeh}

\address{Math Department, Rutgers University,\\
New Brunswick, New Jersey 08854, USA}

%

\begin{abstract}
In this short paper, we review the Dirac equation on the zero-gravity Kerr-Newman spacetime. Our main objective is to provide a correspondence between the classification of the bound states for the zGKN spectrum and the usual hydrogenic states $1s_{1/2}$, $2s_{1/2}$, etc. of the Hydrogen atom.
\end{abstract}

\keywords{Kerr-Newman, Dirac, Hamiltonian, Hydrogen, Point spectrum.}

\bodymatter

\section{The purpose of this paper}

The zero-gravity Kerr-Newman (zGKN) spacetime has been studied extensively\cite{Shadi_zGKN, Shadi_Michael_zGKN_Dyn_System, Shadi_Michael_novel}. In reference \citenum{Shadi_Michael_zGKN_Dyn_System} it was shown that the discrete spectrum of the Dirac Hamiltonian on zGKN is nonempty. In an upcoming paper\cite{zGKN_Spectrum} we classify the discrete spectrum and show that the spectrum is indexed by three integers. See Theorem \ref{main thm} below. It was conjectured\cite{Shadi_Michael_zGKN_Dyn_System, Shadi_Michael_novel} that the discrete spectrum of the Dirac Hamiltonian on zGKN should converge to the Bohr-Sommerfeld spectrum of the usual Hydrogen problem on Minkowski spacetime with a Coulomb potential in the limit as the ring radius of zGKN approaches $0$. This problem remains open but a first step in solving this problem is to determine which states in the zGKN spectrum should correspond to which states in the usual hydrogenic spectrum. For example, which states should correspond to $1s_{1/2}$, $2s_{1/2}$, $2p_{1/2}$, etc.? The purpose of this paper is to provide this correspondence.

\section{The zGKN spacetime}

\subsection{The zGKN spacetime and the Dirac equation}
The zero-gravity Kerr-Newman (zGKN) spacetime\cite{Shadi_Michael_novel} is obtained by formally taking Newton's gravitational constant $G \to 0$ in the Kerr-Newman spacetime. In Boyer-Lindquist coordinates, the resulting spacetime has line element

\begin{equation}\label{zGKN metric}
ds^2_{\bg} = c^2dt^2 - (r^2 + a^2)\sin^2\theta d\varphi^2 - \frac{r^2 + a^2\cos^2\theta}{r^2 + a^2}\big(dr^2 + (r^2 + a^2)d\theta^2\big). 
\end{equation}
The zGKN spacetime is static and its orthogonal slices have the topology of two copies of $\R^3$ glued along a disc in the $z = 0$ plane; this topology is known as the Zipoy topology\cite{Zipoy}. Specifically the spacetime manifold is given by $\R \times \Sigma$ where 
\begin{equation}
\Sigma \,=\, \big\{(r, \theta, \varphi) \mid r \in \R,\, \theta \in [0, \pi], \varphi \in [0, 2\pi) \big\} \setminus R
\end{equation}
where $R$ denotes the ring $R = \{r=0, \theta=\pi/2,\varphi\in[0,2\pi)\}$.

The electromagnetic fields on the Kerr-Newman spacetime $\bF_{\textsc{KN}} =d\bAKN$  do not depend on Newton's gravitational constant and so they survive the $G \to 0$ limit. Hence the zGKN spacetime comes already decorated with the same electromagnetic fields. The four-potential reads
\beq\label{def:AKN}
\bAKN = - \frac{r}{r^2+a^2\cos^2\theta} \left(\textsc{q}dt - \frac{\textsc{q}a}{c} \sin^2\theta\, d\varphi\right).
\eeq

  The field $\bF$ is thus singular on the same ring $\{r=0, \theta=\pi/2,\varphi\in[0,2\pi)\}$ as the metric; for $r$ very large and positive it exhibits an electric monopole of strength $\textsc{q}$ and a magnetic dipole moment of strength $\textsc{q}a$, while
for $r$ very negative it exhibits an electric monopole of strength $-\textsc{q}$ and a magnetic dipole moment of strength 
$-\textsc{q}a$.

Equipped with the Kerr-Newman electromagnetic fields, one can interpret the ring bounding the disc as an elementary particle. With this interpretation, it is natural to consider the Dirac equation
\begin{equation}\label{Dirac eq}
{\tilde\ga}^\mu 
\left(-i\hbar \nabla_\mu  + e A_\mu\right)\Psi + mc^2 \Psi = 0, 
\end{equation}
on the zGKN spacetime and compare its spectral properties to that of the usual Hydrogen problem (i.e. the Dirac equation on Minkowski spacetime with a Coulomb potential centered at the origin in $\R^3$ within the Born-Oppenheimer approximation). 

 In equation (\ref{Dirac eq}), $m$ is the mass of an electron, $-e$ is its fundamental charge, and $A_\mu$ is the 1-form electromagnetic potential for $\mathbf{F}_{\textsc{KN}}$ related via $F_{\mu\nu} = \pd_\mu A_\nu - \pd_\nu A_\mu$. The Dirac matrices $({\tilde\ga^\mu})_{\mu=0}^3$ satisfy $\tilde{\ga}^\mu \tilde{\ga}^\nu + \tilde{\ga}^\nu\tilde{\ga}^\mu = 2 g^{\mu\nu}$, with $g^{\mu\nu}$ the (inverse) metric coefficients of the zGKN metric.

At this point, one should recognize that if one takes the limit $a \to 0$, then (\ref{zGKN metric}) becomes the Minkowski metric, (\ref{def:AKN}) becomes the usual Coulomb potential of a charge $\textsc{q}$ located at the origin, and (\ref{Dirac eq}) becomes the usual Dirac equation on Minkowski spacetime. These statements hold only formally as there are issues with the domain of the Dirac Hamiltonian when trying to make these limits precise. In particular, note that the domain of the $r$ coordinate is $\R$ in zGKN while it's just $(0,\infty)$ for Minkowski spacetime. Nevertheless, the $a \to 0$ limit approach of the usual Hydrogen problem suggests a tantalizing mathematical problem.

\subsection{Separation of variables}

 Using Cartan's frame method\cite{BrillCohen66} with a frame  well-adapted to oblate spheroidal coordinates,
 Chandrasekhar \cite{Chandra76a,Chandra76b}, Page\cite{Page76}, and Toop\cite{Toop76}
 transformed  the Dirac equation (\ref{Dirac eq}) into an equation for a bispinor that allows a clear separation of the $t$, $r$, $\theta$, and $\varphi$ derivatives. The same transformation works for zGKN as well\cite{Shadi_Michael_zGKN_Dyn_System}. By introducing an explicit diagonal matrix $\fD = \fD(r,\theta,\varphi)$ and defining $\hat{\Psi}= \fD^{-1} \Psi$, the Dirac equation becomes
\beq\label{eq:DirSep}
(\hat{R} +\hat{A}) \hat{\Psi} = 0,
\eeq
where $\hat{R}$ and $\hat{A}$ are given by (with $\hbar = c = 1$ units from now on)
\bna
\hat{R}& := & \left(\begin{array}{cccc} imr & 0 &D_- - i e\textsc{q}\frac{r}{\varpi} & 0 \\
0 & imr & 0 & D_+ - ie\textsc{q}\frac{r}{\varpi}\\
D_+ - i e\textsc{q}\frac{r}{\varpi} & 0 & imr & 0 \\
0 & D_- -ie\textsc{q}\frac{r}{\varpi} & 0 & imr \end{array} \right),\\
\hat{A} &:= & \left(\begin{array}{cccc} -m a \cos\theta & 0 & 0 & -L_- \\
0 & -ma\cos\theta & -L_+ &0 \\
0 & L_- & ma\cos\theta & 0 \\
L_+ & 0 & 0 & ma\cos\theta
\end{array}\right),
\ena
where
\beq\label{eq:DpmLpm}
D_\pm := \pm \De \pd_r + \left( \De \pd_t + \frac{a}{\De} \pd_\varphi\right),
\qquad 
L_\pm :=\pd_\theta  \pm i \left( a\sin\theta \pd_t+\csc\theta \pd_\varphi\right),
\eeq
and
\[
\varpi := \sqrt{r^2 + a^2}.
\]
 Once a solution $\hat{\Psi}$ to (\ref{eq:DirSep}) is found, the bispinor $\Psi := \fD\hat{\Psi}$
solves the original Dirac equation (\ref{Dirac eq}).
The explicit form of $\fD$ can be found in reference \citenum{Shadi_Michael_zGKN_Dyn_System}, but it's not needed for this paper.

  Separation of variables is now achieved with the Ansatz that a solution $\hat{\Psi}$ of (\ref{eq:DirSep}) is of the form
\beq\label{chandra-ansatz} 
\hat{\Psi} = e^{-i(Et-\kappa \varphi)} \left( \begin{array}{c}R_1S_1\\ R_2 S_2\\ R_2 S_1\\ R_1 S_2 \end{array}\right),
\eeq
with $E$ a yet to be found energy eigenvalue of the Dirac hamiltonian, $\kappa \in \bbZ + \frac12$, and with
 $R_k$ being complex-valued functions of $r$ alone, and $S_k$ being real-valued functions of $\theta$ alone.  
Let
\beq
\vec{R} := \left(\begin{array}{c} R_1\\ R_2\end{array}\right),\qquad \vec{S} := \left(\begin{array}{c} S_1\\ S_2\end{array}\right).
\eeq
Plugging the Chandrasekhar Ansatz \eqref{chandra-ansatz} into \eqref{eq:DirSep} one easily finds that there must be 
$\la\in\C$ such that 
\beq\label{eq:rad} 
T_{rad}\vec{R} =  E\vec{R},
\eeq
\beq\label{eq:ang}
T_{ang}\vec{S} = \la \vec{S},
\eeq
where
\bna
T_{rad} & :=  \label{eq:Trad} 
& \left(\begin{array}{cc} d_- 
&m\frac{r}{\De} - i\frac{\la}{\De} \\ m\frac{r}{\De}+i\frac{\la}{\De} 
& -d_+ \end{array}\right)
\\
T_{ang}& := \label{eq:Tang} 
& \left(\begin{array}{cc}  ma\cos\theta & l_- \\
 -l_+ &-ma\cos\theta  \end{array}\right)
\ena
The operators $d_\pm$ and $l_\pm$ are ordinary differential operators in $r$ and $\theta$ respectively, 
with coefficients that depend on the unknown $E$, and parameters $a$, $\kappa$, and $e\textsc{q}$:
\bna\label{opdefsA}
d_\pm & := & -i \frac{d}{dr} \pm \frac{-a\kappa +e\textsc{q} r}{\De^2}\\
l_\pm & := & \frac{d}{d\theta} \pm \left( aE\sin\theta - \kappa \csc\theta\right).\label{opdefsB}
\ena
 
 \subsection{The coupled spectral problems for $T_{rad}$ and $T_{ang}$}\label{Eval}
 
 The angular operator $T_{ang}$ in (\ref{eq:ang}) is easily seen to be essentially self-adjoint on
 $(C^\infty_c((0,\pi),\sin\theta d\theta))^2 \subset (L^2((0,\pi),\sin\theta d\theta))^2$, and its self-adjoint extension (also denoted $T_{ang}$) has a
purely discrete spectrum $\la=\la_n(am,aE,\kappa)\in \R$, $n\in \Z\setminus 0$ (see references \citenum{SufFacCos83,WINKLMEIERa}).

With $\la\in\R$ it then follows that the radial operator $T_{rad}$ is essentially self-adjoint on $(C^\infty_c(\R, dr))^2\subset (L^2(\R,dr))^2$;
its self-adjoint extension will also be denoted $T_{rad}$. Moreover, we can take $R_1 = R_2^*$ without loss of generality.\footnote{This follows by multiplying the rows of equation (\ref{eq:rad}) by either $R_1^*$ and $R_2^*$ and adding the equations to conclude that $|R_1| = |R_2|$. Therefore $R_1 = Re^{i\phi_1}$ and $R_2 = Re^{i\phi_2}$. Again, by multiplying by conjugates, one can show that $\frac{d}{dr}(R_1/R_2^*) = 0$ which implies $\frac{d}{dr}(\phi_1 + \phi_2) = 0$.} Thus, we can set
\beq
R_1 =\frac{1}{\sqrt{2}}( u-iv),\qquad R_2  =\frac{1}{\sqrt{2}}( u + iv)
\eeq
 for real funcions $u$ and $v$.
 This brings the radial system (\ref{eq:rad}) into the following standard (Hamiltonian) form
\beq\label{eq:hamil}
(H_{rad} -E)\left(\begin{array}{c} u \\ v \end{array}\right) = \left(\begin{array}{c}0 \\ 0 \end{array}\right),
\eeq
where
\beq\label{eq:Hrad}
H_{rad} := \left(\begin{array}{cc} m \frac{r}{\De} + \frac{\ga r+a\kappa}{\De^2} & -\pd_r + \frac{\la}{\De} \\[20pt]
 \pd_r +\frac{\la}{\De} & -m\frac{r}{\De} + \frac{\ga r+a\kappa}{\De^2}  \end{array}\right),
\eeq
and
\beq
\ga := -e\textsc{q} <0.
\eeq 
Equation (\ref{eq:Hrad}) should be compared with equation (7.105) in reference \citenum{ThallerBOOK}. Specifically, note that as $a \to 0$, $H_{rad}$ approaches the radial Hamiltonian of the usual Hydrogen problem on Minkowski spacetime with a Coulomb potential.

 Using techniques of Weidmann\cite{Weidmann} it is straightforward to show that the essential spectrum of $H_{rad}$ consists of values $E\in (-\infty,1]\cup[1,\infty)$, 
 and its interior is purely absolutely continuous; see   reference \citenum{Shadi_Michael_zGKN_Dyn_System}.
  The remaining task is to characterize the discrete spectrum in the gap, i.e. the eigenvalues $E\in (-1,1)$. 
  In reference \citenum{Shadi_Michael_zGKN_Dyn_System} it was shown that the spectrum is symmetric about $0$, hence it suffices to consider $E>0$.
  
  \subsection{The Pr\"ufer transformed system}
Following reference \citenum{Shadi_Michael_zGKN_Dyn_System}, we transform the equations (\ref{eq:hamil}) and (\ref{eq:ang}) for the four unknowns $(u,v)$ and $(S_1,S_2)$
by defining four new unknowns $(R,\Om)$ and $(S,\Theta)$ via the Pr\"ufer transform 
\beq\label{eq:prufer}
u =\sqrt{2} R \cos\frac{\Om}{2},\quad v = \sqrt{2} R \sin\frac{\Om}{2},\quad S_1 = S \cos\frac{\Theta}{2},\quad S_2 = S \sin\frac{\Theta}{2}.
\eeq
Thus
\beq
R =\half\sqrt{u^2+v^2},\quad\Om = 2\tan^{-1}\frac{v}{u},\quad S = \sqrt{S_1^2+S_2^2},\quad \Theta = 2\tan^{-1}\frac{S_2}{S_1}.
\eeq
As a result, $R_1 =  Re^{-i\Om/2}$ and $R_2 = Re^{i\Om/2}$.  Hence $\hat{\Psi}$ can be re-expressed in terms of the Pr\"ufer variables as
\beq\label{ontology}
\hat{\Psi}(t,r,\theta,\varphi) = R(r)S(\theta)e^{-i(Et-\ka \varphi)} \left(\begin{array}{l}
 \cos(\Theta(\theta)/2)e^{-i\Om(r)/2}\\
\sin(\Theta(\theta)/2) e^{i\Om(r)/2}\\
\cos(\Theta(\theta)/2)e^{i\Om(r)/2}\\
\sin(\Theta(\theta)/2)e^{-i\Om(r)/2}\end{array}\right),
\eeq
and we obtain the following equations for the new unknowns, first
\bna
\frac{d}{dr}\Om    &=& 2 \frac{mr}{\De} \cos\Om + 2\frac{\la}{\De} \sin\Om +2\frac{a\kappa + \gamma r}{\De^2} - 2E ,\label{eq:Om}\\
\frac{d}{dr} \ln R &=& \frac{mr}{\De}\sin\Om - \frac{\la}{\De} \cos\Om ,\label{eq:R}
\ena
and second,
\bna
\frac{d}{d\theta}\Theta &=& -2ma\cos\theta\cos\Theta + 2\left(aE \sin\theta - \frac{\kappa}{\sin\theta}\right)\sin\Theta + 2\la,\label{eq:Theta}\\
\frac{d}{d\theta} \ln S &=& -ma \cos\theta\sin\Theta - \left(aE\sin\theta - \frac{\kappa}{\sin\theta}\right)\cos\Theta. \label{eq:S}
\ena
 Note that the $\Om$-equation (\ref{eq:Om}) is decoupled from $R$, and the $\Theta$-equation (\ref{eq:Theta}) is decoupled from $S$.
  Thus the pair (\ref{eq:Om}), (\ref{eq:Theta}) can be solved together independently of equations (\ref{eq:R}), (\ref{eq:S}), which in turn
  can be integrated subsequently by direct quadrature. 

 We can further simplify the analysis of these systems and reduce the number of parameters involved by noting that 
by defining the constants $a'=ma$, $E'=E/m$, and changing to the variable $r'=mr$, we eliminate $m$ from the system.
 Henceforth we therefore set $m=1$. 


\subsection{Transformation onto a coupled dynamical system on cylinders}
 
  Equations  (\ref{eq:Om}) and (\ref{eq:Theta}) exhibit the independent and the dependent variables explicitly. 
  It is more convenient to transform them to a parametrically coupled pair of autonomous two-dimensional dynamical systems, by introducing a new independent 
variable $\tau$, as follows.
  
 Equation (\ref{eq:Theta}) can be written as a smooth dynamical system in the $(\theta,\Theta)$ plane by introducing 
 $\tau$ such that $\frac{d\theta}{d\tau} = \sin\theta$.
Then, with dot representing differentiation in $\tau$,  we have, 
\beq\label{dynsysTh}
\left\{\begin{array}{rcl}
\dot{\theta} & = & \sin\theta\\
       \dot{\Theta} & = & -2a\sin\theta\cos\theta\cos\Theta+2aE\sin^2\theta\sin\Theta - 2\ka\sin\Theta + 2\la\sin\theta
       \end{array}\right.
\eeq
Identifying the line $\Theta=\pi$ with $\Theta=-\pi$, this becomes a dynamical system on a closed finite cylinder 
$\mc{C}_1:=[0,\pi]\times\Sset^1$. 
 The only equilibrium points of the flow are on the two circular boundaries: 
Two on the left boundary: $S^- = (0,0)$, $N^- = (0,\pi)$; two on the right: $S^+ = (\pi,-\pi)$ and $N^+ = (\pi,0)$.  

For $\ka>0$, the linearization of the flow at the equilibrium points reveals that $S^-$ and $S^+$ are hyperbolic saddle points (with 
eigenvalues $\{1,-2\ka\}$ and $\{-1,2\ka\}$ respectively), while $N^-$ is a  source node (with  eigenvalues 1 and $2\ka$) 
and $N^+$ is a sink node (with eigenvalues $-1$ and $-2\ka$).  The situation with $\ka<0$ is entirely analogous, with the two
critical points on each boundary switching their roles.

Similarly, the $\Om$ equation (\ref{eq:Om}) can  be rewritten as a smooth dynamical system on a cylinder, in this case by setting 
$\tau := \frac{r}{a}$ as new independent variable, as well as introducing a new dependent variable
\beq
\xi := \tan^{-1}\frac{r}{a} = \tan^{-1}\tau 
\eeq
Then, with dot again representing differentiation in $\tau$, (\ref{eq:Om}) is equivalent to
\beq\label{dynsysOm}
\left\{\begin{array}{rcl}
\dot{\xi} & = & \cos^2\xi \\
\dot{\Om} & = & 2a\sin\xi\cos\Om+2\la\cos\xi\sin\Om + 2\ga\sin\xi\cos\xi+2\ka\cos^2\xi - 2aE
\end{array}\right.
\eeq
Once again, identifying $\Om=-\pi$ with $\Om  = \pi$ turns this into a smooth flow on the closed finite cylinder 
$\cC_2 := [-\frac{\pi}{2},\frac{\pi}{2}]\times \Sset^1$.  The only equilibrium points of the flow are on the two circular boundaries. 
For $E\in(0,1)$ there are two equilibria on each: $S^-_E = (-\frac{\pi}{2},-\pi+\cos^{-1}E)$ and $N^-_E=(-\frac{\pi}{2},\pi - \cos^{-1}E)$ 
on the left boundary, and $S^+_E = (\frac{\pi}{2}, -\cos^{-1}E)$ and $N^+_E = (\frac{\pi}{2},\cos^{-1}E)$ on the right boundary.
 $S^\pm_E$ are non-hyperbolic (degenerate) saddle-nodes, with eigenvalues $0$ and $\pm 2a\sqrt{1-E^2}$, while $N^-_E$ is a degenerate 
source-node and $N^+_E$ a degenerate sink-node (see Theorem 2.19(iii) in reference \citenum{QTPDS}).

In reference \citenum{Shadi_Michael_zGKN_Dyn_System}, it was shown that $E$ is an energy eigenvalue of the Dirac Hamiltonian  and the corresponding $\Psi$ is a bound state if and only if there exists a $\la\in \R$ such that each of the two dynamical systems above possesses a {\em saddles connector,} i.e. an orbit on $\mathcal{C}_1$ connecting the two saddle-nodes $S^-$ and $S^+$ in the $\Theta$-system (\ref{dynsysTh}) 
and an orbit on $\mathcal{C}_2$ connecting the two saddle-nodes $S^-_E$ and $S^+_E$ in the $\Omega$-system (\ref{dynsysOm}). 

Given a dynamical system on a cylinder, there corresponds an integer known as the \emph{winding number} which describes how many times an orbit in the dynamical system winds around the cylinder before terminating at an equilibrium point. See reference \citenum{Shadi_Michael_zGKN_Dyn_System} and the upcoming paper \citenum{zGKN_Spectrum}. 
For the $\Omega$ system, saddles connectors with different winding numbers correspond to different energy values (with energy increasing as the winding number increases). In reference \citenum{Shadi_Michael_zGKN_Dyn_System} it was shown that a bound state $\Psi$ exists corresponding to winding number $N_\Theta = 0$ for the $\Theta$ system and winding number $N_\Omega = 0$ for the $\Omega$ system (and $\kappa = \half$). In the upcoming paper \citenum{zGKN_Spectrum}, we improve on this result by fully classifying the spectrum. Specifically, we prove the following theorem. 

\medskip

\begin{theorem}\label{main thm}\:
Set $a_{\rm{max}} = 1 - \frac{1}{\sqrt{2}}$ and $\gamma_{\rm{min}} = -\half$. Fix $a \in (0, a_{\rm{max}})$, $\gamma \in (\gamma_{\rm{min}}, 0)$, and $\kappa \in \Z +\half$. Assume $\Psi$ is of the form \emph{(\ref{ontology})} constructed from solutions of \emph{(\ref{eq:Om})} - \emph{(\ref{eq:S})}.

\begin{itemize}
\item[$\bullet$] Suppose $N_\Theta \geq 0$ is an integer. For all integers $N_\Omega \geq 0$, there is a bound state $\Psi$ such that the $\Theta$ system and $\Omega$ system have winding numbers $N_\Theta$ and $N_\Omega$, respectively. There are no bound states with $N_\Omega \leq - 1$. 

\item[$\bullet$] Suppose $N_\Theta \leq -1$ is an integer. For all integers $N_\Omega \geq 1$, there is a bound state $\Psi$ such that the $\Theta$ system and $\Omega$ system have winding numbers $N_\Theta$ and $N_\Omega$, respectively. There are no bound states with $N_\Omega \leq 0$.
\end{itemize}
\end{theorem}

\medskip

Our conditions on $a$ and $\gamma$ are used to ensure that there are no bound states with $N_\Omega \leq -1$ for $N_\Theta \geq 0$ and no bound states with $N_\Omega \leq 0$ 
 for $N_\Theta \leq -1$, which, as we will see in the next section, occurs in an analogous way for the familiar hydrogenic Dirac operator on a Minkowski background with a Coulomb potential. 
 
When restoring units and mass $m$, we have $a_{\rm{max}} = (1 - \frac{1}{\sqrt{2}})\frac{\hbar}{mc}$ and $\gamma_{\text{min}} = -\half \hbar c.$ 
 If, with the hydrogenic problem in mind, we set  
 $\gamma = -Ze^2$, then $\gamma \in (\gamma_{\rm min}, 0)$ implies $\frac{Ze^2}{\hbar c} < \frac{1}{2}$, that is, $Z < \frac{137.036}{2}$.  Now let's compare these conditions with the conditions on $Z$ for the familiar hydrogenic Dirac operator on a Minkowski background with a Coulomb potential, for which
essential self-adjointness breaks down for $Z > 118$. For the Dirac operator on zGKN, there is no condition for essential self-adjointness\cite{Shadi_Michael_zGKN_Dyn_System}.
So our  conditions are probably not optimal.

\section{Relating the zGKN bound states to the usual hydrogenic states}

In this section we relate the zGKN bound states found in Theorem \ref{main thm} to the usual hydrogenic states of the Dirac problem on Minkowski spacetime with a Coulomb potential. The main objective is to relate the winding numbers which appear in Theorem \ref{main thm} to the usual spectroscopic notation $n\ell_j$ of hydrogenic states.

\subsection{The correspondence between winding numbers and the usual spectroscopic notation of hydrogenic states}

Based on the above results, for fixed $a$ and $\ga$, the discrete spectrum of our Dirac Hamiltonian is indexed by {\em three} integers: $N_\Omega$, $N_\Theta$,  and $2\kappa$.  
By contrast, the energy spectrum of special relativistic Hydrogen, i.e., the Dirac Hamiltonian for a point-like electron in ordinary Minkowski space interacting with a Coulomb point charge at the origin, is indexed by {\em two} integers only, namely the main (or Bohr's) quantum number, often denoted by $n$, and the spin-orbit quantum number\footnote{The spin-orbit quantum number is called $\kappa_j$ in reference \cite{ThallerBOOK}; it should not be confused with our $\kappa$, which is the eigenvalue of the $z$-component of angular momentum, for which Thaller uses the notation $m_j$.}, i.e., the set of eigenvalues of the spin-orbit operator $K = \beta(2\bS\cdot\bL +1)$.

In the limit $a \to 0$, the angular Hamiltonian (\ref{eq:Tang}),  takes the simple form
\beq\label{eq:Tang0}
\fa_\ka := \lim_{a\to 0} T_{ang} =  i\si_2 \pd_\theta + \frac{\ka}{\sin\theta} \si_1.
\eeq
From reference \citenum{WINKLMEIERa}, if $\la$ is an eigenvalue of $T_{ang}$, then $k := \lim_{a \to 0} \la$ is an eigenvalue of $\fa_\ka$, also the limit defining $k$ exists since $\la$ is analytic in $a$. Note that $\fa_\ka$ is independent of $E$ unlike $T_{ang}$.

In the limit $a\to 0$, the formal limit of the radial Hamiltonian (\ref{eq:Hrad}) coincides with the radial Hamiltonian arising in the special relativistic Hydrogen problem (e.g. see reference \citenum{ThallerBOOK}, eq. (7.105)):
\beq\label{eq:Hrad0}
\fh_{k} := \lim_{a\to 0} H_{rad} = \left(\begin{array}{cc} m  + \frac{\gamma}{r} & -\pd_r + \frac{k}{r} \\[20pt]
 \pd_r +\frac{k}{r} & -m + \frac{\gamma}{r}  \end{array}\right).
\eeq
Therefore $k$ can be identified with the \emph{spin-orbit coupling}.

The spectrum of $\fa_\ka$ is completely understood\cite{WINKLMEIERa}. In particular, for all half-integers $\ka \in Z + \half$, the operator $\fa_\ka$ is essentially self-adjoint and has a discrete spectrum indexed by a nonzero integer which we call $N$:

\beq\label{kintermsofN}
k = -\sgn(N) \left( |N| + |\ka| - \half\right),
\eeq
as well as a complete set of eigenvectors $\vec{S}_{N,\ka}$ that are explicitly known and can be expressed in terms of Jacobi polynomials.
\begin{equation}
\vec{S}_{N,\ka}(\theta) := \sin^{\ka+\half} \theta \left(\begin{array}{c} 
-\sqrt{\cot\frac{\theta}{2}} P_{|N|-1}^{\ka-\half,\ka+\half}(\cos\theta) \\
\sgn(N)\sqrt{\tan\frac{\theta}{2}} P_{|N|-1}^{\ka+\half,\ka-\half}(\cos\theta)
\end{array}\right).
\end{equation}
Note that the above holds for $\ka>0$.  To find the eigenvectors for $\ka<0$, recognize that if $\vec{S}$ is an eigenvector of $\fa_\ka$, then $i\si_2 \vec{S}$ is an eigenvector of $\fa_{-\ka}$. 

From the above and the definition of $\Theta$ it follows that the saddles connectors of the $\Theta$-system that correspond to the eigenvectors of the angular Hamiltonian are given explicitly by the formula
\begin{equation}
\Theta_{N,\ka}(\theta) =  -\sgn(N) \left\{ 2\tan^{-1}\left(  \frac{P_{|N|-1}^{|\ka|+\half,|\ka|-\half}(\cos\theta)}{P_{|N|-1}^{|\ka|-\half,|\ka|+\half}(\cos\theta)}\tan\frac{\theta}{2}\right) - \pi \boldsymbol{\chi}_{\ka<0}\right\}.
\end{equation}
Here the branch of $\tan^{-1}$ needs to be chosen in such a way that $\Theta_{N,\ka}$ is continuous on $[0,\pi]$.  

The above formula implies that we can fix the initial value of the saddles connectors to be
\begin{equation}
\Theta_{N,\ka}(0) = \left\{\begin{array}{cc} 0 & \ka>0\\ \sgn(N)\pi & \ka<0.
\end{array}\right.
\end{equation}
and  that for the final value of those connectors, we need
\begin{equation}
\Theta_{N,\ka}(\pi) = \left\{\begin{array}{cc} -\sgn(N)\pi & \ka>0 \\ 0 & \ka<0 \end{array}\right.\qquad [\mbox{mod} 2\pi]
\end{equation}
in agreement with what we have already observed about the boundary values of the saddles connectors for the $\Theta$-system (\ref{dynsysTh}).

 Furthermore, from the properties of Jacobi polynomials, it follows that
for $\Theta_{N,\ka}$ to be a continuous function of $\theta$ on $[0,\pi]$, we need
\beq\label{Thetafinal}
\Theta_{N,\ka}(\pi) = -2\pi\sgn(N)(|N|-1)+\left\{\begin{array}{cc} -\sgn(N)\pi & \ka>0 \\ 0 & \ka<0 \end{array}\right.
\eeq
Thus we establish a correspondence between the integer $N$ and the winding number $N_\Theta$ of the $\Theta$-saddles connectors in the case $a=0$.  
\beq\label{Thetaconn}
N_\Theta \sim 
\left\{\begin{array}{cc} N - 1 & N \geq 1 
\\ 
N & N \leq -1, \end{array}\right.
\eeq
where `$\sim$' simply means a correspondence. 

 From the relationship (\ref{kintermsofN}) between the eigenvalue $k$ and the number $N$ it follows that instead of $N$ we can equally well label the angular eigenstates by the integer $k$, in which case
\begin{equation}
N \,=\, -\sgn(k) \left( |k| - |\ka| + \half\right).
\end{equation}

We now turn our attention to the radial Hamiltonian $\fh_k =\lim_{a\to 0} H_{rad}$. This operator is closely related to the radial Hamiltonian $h_k$ of the special-relativistic Hydrogen problem, as formulated by Dirac\cite{Dir28}:
\begin{equation}
h_k = -i\si_2 \pd_r + m \si_3 + \frac{k}{r} \si_1 + \frac{\ga}{r} I_{2\times 2}.
\end{equation}

The only difference between $h_k$ and $\fh_k$ is their domains:  Since $H_{rad}$ is defined on the double-sheeted Sommerfeld space, this is inherited by its $a\to 0$ limit $\fh_k$, which is still defined on two copies of Minkowski space glued together along a timelike line.  In particular the $r$ variable in $\fh_k$ has the range $(-\infty,\infty)$.  By contrast, the $r$ variable in $h_k$ goes from $0$ to $\infty$.
  The eigenvalue problem for $h_k$ was shown to be exactly solvable by Gordon \cite{Gor28}, who proved the discrete spectrum to coincide exactly with the Bohr-Sommerfeld spectrum, and found explicit formulas for the eigenfunctions in terms of generalized Laguerre polynomials.

 Consider first the restriction of $\fh_k$ to functions supported in the sheet $r > 0$.   As described in detail in section 7.4 of Thaller's book\cite{ThallerBOOK}, this operator, for all $k \in \Z\setminus\{0\}$, is essentially self-adjoint on $C^\infty_c((0,\infty))$, has a discrete spectrum in $(0,1)$ and a complete set of eigenvectors,  if $-\sqrt{3}/2 < \ga < 0$.  The discrete spectrum is indexed by two integers, $n\geq 1$ and $k = -n,\dots,-1,1,\dots,n-1$.  Let $M:= n - |k|$.  Then we have
\begin{equation}
E_{M,k} \,=\, \frac{m}{\sqrt{ 1+\left( \frac{\ga^2}{M + \sqrt{ k^2 - \ga^2}}\right)^2}}
\end{equation}
Note that the case $k > 0$ and $M = 0$ is excluded. 

Gordon \cite{Gor28} computed the corresponding eigenfunctions in terms of generalized Laguerre polynomials:  Let $(\phi_1,\phi_2)$ be defined by 
\begin{equation}
u = \sqrt{1+E_{M,k}}(\phi_1+\phi_2),\qquad v = \sqrt{1-E_{M,k}}(\phi_1-\phi_2)
\end{equation}
where $(u,v)$ is an eigenfunction of $\fh_k$ with eigenvalue $E_{M,k}$ (and we have set $m=1$).  Then, using the abbreviations
\begin{equation}
\rho := \sqrt{k^2 - \ga^2},\qquad \eta := \sqrt{1-E_{M,k}^2}
\end{equation}
 we have that for all nonnegative integers $M$ and real constants $c_1,c_2$ such that
\begin{equation}
\mu := \frac{c_1}{c_2} = \frac{M}{k + \frac{\ga}{\eta}}.
\end{equation}
we have
\begin{align}
\phi_1(r) \,&=\, c_1 e^{-\eta r} r^\rho F(-M+1, 2\rho+1, 2\eta r)\\
\phi_2(r) \,&=\, c_2 e^{-\eta r} r^\rho F(-M,2\rho+1,2\eta r)
\end{align}
where $F$ denotes Gauss\rq{}s {\em confluent hypergeometric function}
\begin{equation}
F(\al,\beta,x) \,=\, 1 + \frac{\al}{1!\beta}x+\frac{\al(\al+1)}{2!\beta(\beta+1)}x^2 + \dots
\end{equation}
Note that when $\al$ is a negative integer, the above series terminates, and $F$ is a polynomial of degree $-\al$, which (up to a numerical factor) is the generalized Laguerre polynomial $L_{-\al}^{(\beta-1)}(x)$.  

Accordingly, since $\Om = 2 \tan^{-1}\frac{v}{u}$, the corresponding solution to the $a \to 0$ limit of the $\Om$-equation (\ref{eq:Om}) will be
\beq\label{Omsol}
\Om(r) = 2 \tan^{-1}\left( \sqrt{\frac{1-E}{1+E}}\ \frac{\mu F(-M+1,2\rho+1,2\eta r) - F(-M,2\rho+1,2\eta r)}{\mu F(-M+1,2\rho+1,2\eta r) + F(-M,2\rho+1,2\eta r)}\right)
\eeq
The values of $\Om$ at $r=0$ and $r=\infty$ can thus be calculated (modulo $2\pi$):
\begin{align}
\Om(0) \,&=\, 2 \tan^{-1}\left( \sqrt{\frac{1-E}{1+E}}\ \frac{\mu-1}{\mu+1}\right) \,=  \sin^{-1}(\frac{-\ga}{k})
\\
 \Om(\infty) \,&=\, -2\tan^{-1}\left( \sqrt{\frac{1-E}{1+E}}\right) \,=\, -\cos^{-1}E
\end{align}
 which is in agreement with the analysis of the equilibrium points of the corresponding dynamical system.  We choose our principal branch of $\sin^{-1}$ depending on the sign of $k$:
\begin{align}
\Om(0) \,&=\, \left\{\begin{array}{ll} 
\sin^{-1}(\frac{-\ga}{k}) & \mbox{ if }k<0.
\\
-\pi - \sin^{-1}(\frac{-\ga}{k}) & \mbox{ if }k>0
\end{array}\right.
\end{align}
On the other hand, from the properties of Laguerre polynomials, it follows that the denominator of the rational function in (\ref{Omsol}) has $M$ zeros in $[0, \infty)$ when $k < 0$ and $M-1$ zeros when $k>0$ (note that this explains why $k \leq n -1$). Thus in order for $\Om$ to be a continuous function of $r$, the branch of $\tan^{-1}$ needs to be chosen such that $-2\pi $ gets added to the value every time $r$ crosses one of those poles, which implies that
\begin{equation}
\Omega(\infty) \,=\, -2\pi M - \cos^{-1}E
\end{equation}
which holds for both $k < 0$ and $k > 0$ (this is why we chose $\Omega(0) = -\pi - \sin^{-1}(\frac{-\ga}{k})$ for $k > 0$). Thus we establish the following correspondence between the integer $M$ and the winding number $N_\Omega$ for the $\Omega$-saddles connectors in the case $a = 0$. 
\beq\label{Om Conn}
N_\Omega \,\sim\, M. 
\eeq

Now consider the case $a > 0$ (i.e. the Dirac Hamiltonian on zGKN). The bound states are indexed by three integers $N_\Theta, N_\Omega$, and $2\kappa$ appearing in Theorem \ref{main thm}. Using (\ref{Om Conn}) and (\ref{Thetaconn}), we can define a correspondence between the integers $N_\Theta, N_\Omega, 2\kappa$ and the usual spectroscopic notation $n\ell_j$ of hydrogenic states. Given $N_\Theta$ and $N_\Omega$, define $N$ and $M$ via  
\beq
N \,:=\,
\left\{\begin{array}{cc} N_\Theta + 1 & N_\Theta \geq 0 
\\ 
N_\Theta & N_\Theta \leq -1, \end{array}\right. \quad \text{ and } \quad M \,:=\, N_\Omega. 
\eeq
Then $n$, $\ell$, and $j$ appearing in $n\ell_j$ are given by 
\begin{align}
k \,&:=\, -N - \sgn(N)\big(|\kappa| - \half\big)
\\
n\,&:=\, M + |k|
\\
j\,&:=\, |k| - \half
\\
\ell \,&:=\, j + \sgn(k)\half
\\
m_j \,&:=\, \kappa. 
\end{align}

This establishes the desired correspondence between the winding numbers appearing in Theorem \ref{main thm} and the usual spectroscopic notation $n \ell_j$ of hydrogenic states. For example, suppose we ask: which state in zGKN corresponds to $2p_{1/2}$ with $m_{1/2} = -\half$? Well $\ell = 1$ implies $1 = \half + \sgn(k)\half$. Therefore $k$ is positive and so $j = \half$ implies $k = 1$. Therefore $N = -1$ implies $N_\Theta = -1$. Lastly, $n = 2$ implies $M = 1$ and so $N_\Omega = 1$. Thus  the state $2p_{1/2}$ with $m_{1/2} = - \half$ corresponds to $N_\Theta = -1$, $N_\Omega = 1$, and $\kappa = -\half$.

\subsection{Breaking degeneracies}

It is well known that the energy levels of the special-relativistic Hydrogen Hamiltonian are independent of $m_j$, the eigenvalue of the $z$-component of total angular momentum, while the eigenstates of it do depend on $m_j$.  As a result, all energy levels are degenerate in that case, i.e. have multiplicity at least two.  
In the case of our Hamiltonian $H_{rad}$, this degeneracy is broken by the appearance of $a \kappa$ in equation (\ref{eq:Hrad}). This yields a hyperfine-like splitting of the spectral lines, which in the standard setting is thought of as being a consequence of the nucleus having a magnetic dipole moment.  In our setting, it can be thought of as following from the nonzero $d\varphi$ term in equation (\ref{def:AKN}) which produces a magnetic dipole moment.

In addition to the above symmetry-related degeneracy, Dirac's original model for special-relativistic Hydrogen also has {\em accidental} degeneracies; the energy eigenvalues do not depend on the {\em sign} of the spin-orbit quantum number $k$, only on its magnitude, while the eigenstates do depend on the sign of $k$.  The celebrated  experiment of Lamb \cite{Lam55} showed that this degeneracy is only in Dirac's model, not in nature, i.e., the measured energy levels of orbitals corresponding to $k$ and $-k$ are slightly different.  This difference is known as the {\em Lamb shift}.  In our model, these accidental degeneracies are also broken producing a small but observable Lamb shift-like effect. 

Recall that hyperfine splitting and the Lamb shift are calculated using perturbative techniques. It is important to note that our model does not use perturbation theory to calculate energy differences.    

In the upcoming paper\cite{zGKN_Spectrum}, we will report on numerical results of the zGKN spectrum showing the breaking of these degeneracies for various values of $a$.

\medskip
\medskip

\newpage

\bibliographystyle{ws-procs961x669}


\end{document}